\documentclass[letterpaper]{article}
\usepackage{aaai}

\usepackage{times}
\usepackage{helvet}
\usepackage{courier}
\usepackage{mathptmx}
\usepackage{enumitem}
\setlist[itemize]{noitemsep, topsep=0pt}
\usepackage[utf8]{inputenc}
\usepackage{graphicx}
\usepackage[final]{listings}
\usepackage{float}
\usepackage{tikz}
\usetikzlibrary{calc}
\usepackage{url}
\usepackage{multirow}
\usepackage{multicol}
\usepackage{booktabs}

\usepackage[final,commentmarkup=todo,authormarkup=brackets,highlightmarkup=uwave]{changes}
\definechangesauthor{FC}
\definechangesauthor{MRV}

\usepackage{amsmath,amssymb}
\usepackage{ulem}
\newcommand{\audioImagePath}[1]{images/waveforms/audio#1.pdf}
\newcommand{\audioImage}[1]{\includegraphics[height=10pt, width=10pt]{\audioImagePath{#1}}}

\usepackage{todonotes}
\presetkeys%
    {todonotes}%
    {inline,backgroundcolor=gray}{}

\restylefloat{figure}
\restylefloat{table}

\newfloat{lstfloat}{htb}{lop}
\floatname{lstfloat}{Listing}

\usetikzlibrary{positioning,calc,fit,decorations.pathreplacing}

\begin{document}
\title{Using DeepProbLog to perform Complex Event Processing on an Audio Stream}
\author{Marc Roig Vilamala\textsuperscript{1,*}, Tianwei Xing\textsuperscript{2}, Harrison Taylor\textsuperscript{1}, Luis Garcia\textsuperscript{2}, Mani Srivastava\textsuperscript{2}, \\
\textbf{\Large Lance Kaplan\textsuperscript{3}, Alun Preece\textsuperscript{1}, Angelika Kimmig\textsuperscript{4}, Federico Cerutti\textsuperscript{1,5}} \\
\textsuperscript{1}{Cardiff University}\\
\textsuperscript{2}{University of California, Los Angeles}\\
\textsuperscript{3}{Army Research Laboratory}\\
\textsuperscript{4}{KU Leuven, Department of Computer Science; Leuven.AI}\\
\textsuperscript{5}{University of Brescia}\\
\textsuperscript{*}{Corresponding author: RoigVilamalaM@cardiff.ac.uk}\\
}

\maketitle

% \author[cardiff]{Marc Roig Vilamala \corref{cor1}}
% \ead{RoigVilmalaM@cardiff.ac.uk}

% \author[ucla]{Tianwei Xing}

% \author[cardiff]{Harrison Taylor}

% \author[ucla]{Luis Garcia}

% \author[ucla]{Mani Srivastava}

% \author[arl]{Lance Kaplan}

% \author[cardiff]{Alun Preece}

% \author[kuleuven]{Angelika Kimmig}

% \author[cardiff,brescia]{Federico Cerutti}

% \cortext[cor1]{Corresponding author.}
% \address[cardiff]{Cardiff University}
% \address[ucla]{University of California, Los Angeles}
% \address[arl]{Army Research Laboratory}
% \address[kuleuven]{KU Leuven, Department of Computer Science; Leuven.AI}
% \address[brescia]{University of Brescia}

\begin{abstract}
    In this paper, we present an approach to Complex Event Processing (CEP) that is based on DeepProbLog. This approach has the following objectives: (i) allowing the use of subsymbolic data as an input, (ii) retaining the flexibility and modularity on the definitions of complex event rules, (iii) allowing the system to be trained in an end-to-end manner and (iv) being robust against noisily labelled data.
    Our approach makes use of DeepProbLog to create a neuro-symbolic architecture that combines a neural network to process the subsymbolic data with a probabilistic logic layer to allow the user to define the rules for the complex events.
    We demonstrate that our approach is capable of detecting complex events from an audio stream. We also demonstrate that our approach is capable of training even with a dataset that has a moderate proportion of noisy data.
\end{abstract}

% \begin{keyword}
%     Complex Event Processing \sep Neuro-symbolic architecture \sep Learning with sparse data \sep Robustness
% \end{keyword}

%\tableofcontents

%\comment[id=MRV]{We are intending to submit to ICLP 2020. Regular papers must not exceed 14 pages including bibliography. Regular papers may be supplemented with appendices for proofs and details of datasets which do not count towards the page limit and which will not be made available as appendices to the published paper.

%Current deadlines are \sout{8th of May for abstracts and 15th of May for paper submission} 15th of May for abstracts and \sout{22nd of May} 5th of June for paper submission.}

%%%%%%%%%%%%%%%%%%%%%%%%%%%%%%%%%%%%%%%%%%%%%%%%%%%%%%%%%%%%%%%%%%%%%%%%%%%%%%%%%
\section{Introduction}
\label{sec:introduction}

Complex Event Processing (CEP) systems process data streams and detect situations of interest, or \textit{complex events}, which aggregate atomic \textit{events}, or \textit{simple events}. CEP systems detect spatio-temporal relationships between sets of simple events, which form complex events.
CEP systems have been applied in many different areas, such as business activity monitoring \cite{StockMarket2012}, sensor networks \cite{Anicic2012_2} and weather reports \cite{Anicic2012}. Most CEP approaches allow the user to define rules which express the conditions under which a complex event occurs. Then, the CEP system uses those rules to detect when those circumstances happen in the given stream of input data. However, defining rules over raw streams of data can be challenging. For example, it is not feasible to define rules directly over raw images, audios or videos. %This is because, while technically possible, it is not feasible to manually define rules to identify which object appears in an image. As a result, most CEP approaches are limited in which types of data they can use. 
In this paper, we will refer to these types of data for which we cannot (easily) manually define rules to extract the information we want as \emph{subsymbolic data}.

Some new CEP approaches \cite{ROLDAN2020113251,Fusion} have been created to incorporate the use of subsymbolic data. However, as we will discuss in Section \ref{sec:critical_analysis}, they require pre-trained neural networks to work, which are not always available. While it is possible to train these neural networks separately, it can be costly to obtain training data for that case. As such, we want an approach that can train in an end-to-end manner. This means that we want a system that can be trained using only labels for the complex events. While some approaches already allow for such end-to-end training \cite{Neuroplex}, they significantly limit the flexibility and modularity offered when defining the rules for the complex events. This makes it difficult, or even impossible, for the user to precisely express the conditions under which a complex event occurs, particularly for the more complex situations in which it may happen. 
In this paper, we aim to propose an approach to CEP that can be trained to use new types of subsymbolic data without limiting the flexibility and modularity of the rule definitions. 

Due to the difficulty of labelling complex events, it is also possible that the training dataset will contain a portion of incorrectly labelled complex events. As such, another aim for our approach is to be robust against situations where a moderate portion of the training data has incorrect labels. For this paper, we will refer to the incorrectly labelled data in our training dataset as noisy data. This type of error could occur either due to a genuine mistake by the person labelling the dataset or due to malicious intent.
%Another aim for our approach was to make it robust against adversarial conditions, according to current AI ethical principles \cite{board2019ai}. These indicate that AI systems should be reliable, meaning that the systems should behave as expected even in sub-optimal conditions. For this paper, we will be focusing on situations in which the training data is noisy, meaning that part of the labels in the training dataset are incorrect.

As such, we wanted to create an AI system that is capable of performing CEP while fulfilling the following objectives:

\begin{enumerate}
    \item Being able to operate on subsymbolic data streams.

    \item Retaining flexibility and modularity in rule definitions.
    
    \item Being able to perform end-to-end training.
    
    \item Being robust against noisily labelled data.
\end{enumerate}

Currently, none of the approaches to this type of problem cover all such objectives. In Section \ref{sec:critical_analysis} we will explain the limitations of existing approaches.

%None of the existing approaches cover all such objectives. For instance, Neuroplex \cite{Neuroplex} employs a neuro-symbolic architecture that allows it to use images and audio as input while requiring only small amounts of training data. However, in order to archive that, it significantly limits the flexibility when defining the complex event rules, which goes against our second objective. On the other hand, \cite{ROLDAN2020113251,Fusion} present approaches that retain the flexibility of the rule definitions while being able to operate on subsymbolic data. However, they require pre-trained neural networks, for which it would be costly to obtain training data. Existing approaches and their limitations are further explored in Section \ref{sec:critical_analysis}.

In this paper, we propose an approach based on DeepProbLog \cite{deepproblognips,MANHAEVE2021103504} to detect complex events from an audio stream. DeepProbLog allows us to combine a neural network with probabilistic logic rule definitions. As such, the neural network can be used to process the subsymbolic data, which can then be used within the probabilistic logic to detect the patterns that form complex events. Furthermore, the probabilistic logic allows users to easily define the rules for the complex events. DeepProbLog also allows us to train the system in an end-to-end manner, thus fulfilling the first three objectives. For a background explanation of DeepProbLog, see Section \ref{sec:tools_used}. Meanwhile, Section \ref{sec:methodology} explains how we have used DeepProbLog to perform complex event detection.

In order to evaluate the performance of our approach we have generated synthetic datasets. For more details on the dataset generation, see Section \ref{sec:generating_datasets}. Then, in Section \ref{sec:results} we evaluate the performance of our approach after training with the generated datasts. First, we demonstrate that our approach is capable of detecting complex events from an audio stream. Then, we also evaluate how robust our approach is against noisy datasets in the context of incorrectly labelled data. In order to evaluate this, we have generated datasets with a percentage of data that has been incorrectly labelled in a random manner. We have generated datasets with different percentages of incorrectly labelled data in order to evaluate how robust our approach is to this type of noise. 
%a poisoning attack on the training data. This attack consists of changing a percentage of labels for the training data into a random label. 
%We will refer to the portion of training labels that have been changed as the noise in the dataset. 
%We have generated datasets with different percentages of poisoned data in order to evaluate up to which level our approach still trains correctly. 
As we show in Section \ref{sec:results:random_noise}, our approach is robust against moderate amounts of noisy data, with an almost imperceptible decrease in performance after training with a dataset where 20\% of the training data is noise. However, higher percentages of noisy data lead to unreliable results, as the system is not able to consistently train correctly.

Finally, in Section \ref{sec:conclusion} we provide final conclusions on the results and discuss potential areas for future research.

\section{Background}
\label{sec:tools_used}

In this section, we provide background information on complex event processing (CEP). We also give a general overview of ProbLog and DeepProbLob, which are used in our approach. 

\subsection{Complex event processing}

Complex Event Processing (CEP) systems aim to identify aggregations of events that form complex events. Following \cite{Luckham2002}, an \emph{event} is an
object that can be subjected to computer processing and it signifies, or
is a record of, an activity that has happened. For instance, a record of a temperature reading or the value of a stock at a certain point in time can be considered events. However, for this paper, we will be focusing on events that come in the form of subsymbolic data and, more specifically, audio.
CEP is used to automatically detect situations of interest for the user. For instance, in an emergency response setting, CEP could be used to detect rioting in the streets by detecting a combination of people shouting, glass shattering and sirens. In this context, the riot would be the complex event while the individual sounds would be the events that form it. An expert would be required in order to define which combinations of sounds form which complex events.
%In this context, a song could be considered a complex event, as it is an aggregation of events (i.e. intro, chorus, outro, etc.). Although we will not explore it in this paper, it is also worth considering that most of these systems are hierarchical. For instance, a chorus is the aggregation of a set of musical notes and lyrics. At the same time, songs can be aggregated into an album. CEP systems are used to detect certain pre-determined combinations of events that form the type of complex event that the user is interested in. For instance, detecting a certain combination of instruments or notes could indicate to the user that a specific genre of music is being played.

In general, an \emph{event} has
three main aspects:

\begin{itemize}
\item
  \emph{Form}: the form of an event is an object with particular
  attributes or data components, for instance the time period of the
  activity;
\item
  \emph{Significance}: an event signifies an activity, hence an event's
  form usually contains data describing the activity it signifies;
\item
  \emph{Relativity}: an activity is related to other activities. Events
  have the same relationships to one another as the activities they
  signify. The \emph{relativity} of an event refers to the set of relationships between that event and other events.
  An event's form usually encodes
  its relativities, i.e., methods to reconstruct the relationships with
  other events.
\end{itemize}

It is therefore important to notice that an event is not just a message
or a record of an activity: the forms of events may be messages, but the
events also have significance and relativity. In particular, the three
main partial, transitive, and antisymmetric relationships between events
are:

\begin{itemize}
\item
  \emph{Time}: a relationship that orders events. %While it relies on
%   timestamps, in general we need to be open to the possibility of
%   uncertainty associated to the relationships between events: clocks
%   timestamping events might not be related between them.
\item
  \emph{Cause}: a dependence relationship between activities. An
  activity (event) depends upon other activities (events) if it happened
  only because the other activities (events) happened. If event \(B\)
  depends upon event \(A\), then \(A\) \emph{caused} \(B\). If neither
  caused the other, they are \emph{independent}.\footnote{This
    computational notion of causality is ostensibly more limited than
    the notion of causality in philosophy and science in general: an
    interested reader is referred to \cite{pearl2009}}
\item
  \emph{Aggregation}: if event \(A\) signifies an activity that consists
  of the activities of a set of events \(B_{1},B_{2},\ldots,\ B_{n}\),
  then \(A\) is an \emph{aggregation} of all the events \(B_{i}\).
  Conversely, \(B_{i}\) are \emph{members} of \(A\). Aggregation is an
  abstraction relationship: usually event \(A\) is created when a set of
  events \(\{ B_{i}\}\) happens. \(A\) is a higher-level event and we
  call it a \emph{complex event}. \(A\)'s members are the events that
  \emph{caused} it. Aggregation can be referred to also as
  \emph{vertical causality}.
\end{itemize}

In the context of this paper, CEP aims at identifying such aggregation rules, so to make the
activities in a complex system understandable to humans. 
More specifically, we will be splitting the input audio into short segments of audio (1 second long). Each of those segments will be considered an event. Each of event will have a timestamp attached to them, which will indicate at what time the corresponding segment of audio started. For simplicity, in this paper we will be aggregating events into a complex event if the same type of sound (based on a set of pre-defined classes) repeats within a certain window of time. We will evaluate our approach with window sizes between 2 and 5 seconds.
%CEP has been applied in many different areas, including finance \cite{StockMarket2012}, transport \cite{ZAPPIA201210408}, machine-to-machine communication \cite{BRUNS20151235}, weather monitoring \cite{Anicic2012} and detecting IoT security attacks \cite{ROLDAN2020113251}, among many others. 

\subsection{ProbLog}

ProbLog \cite{DeRaedt2007} is a probabilistic logic programming language. ProbLog allows users to encode complex interactions between different components. A ProbLog program consists of a set of probabilistic facts $F$ and a set of rules $R$. Facts have the form $p::f$ where $p$ is a value between 0 and 1, representing the likelihood of the fact being true, and $f$ is an atom. Atoms are expressions of the form $q(t_1, ..., t_n)$ where $q$ is a predicate and $t_i$ are terms. Rules have the form $h~{:}{-}~b_1,...,b_n$ where $h$ is an atom and $b_i$ are literals. A literal is an atom or the negation of an atom.

One convenient syntactic extension is an annotated disjunction (AD), which is an expression of the form $
p_1 :: h_1; ... ; p_n :: h_n~{:}{-}~b_1, ..., b_m.
$ where the $p_i$ are probabilities so that $\sum p_i = 1$, and $h_i$ and $b_j$ are atoms.
The meaning of an AD is that whenever all $b_i$ hold, $h_j$ will be true with probability $p_j$, with all other $h_i$ false (unless other parts of the program make them true). This is convenient to model choices between different categorical variables.
ProbLog programs with annotated disjunctions can be transformed into equivalent ProbLog programs without annotated disjunctions  \cite{deraedt15}. 

\subsection{DeepProbLog}
\label{sec:tools:deepProbLog}

DeepProbLog \cite{deepproblognips,MANHAEVE2021103504} is a neural probabilistic logic programming language that allows the user to create neuro-symbolic architectures. DeepProbLog allows the user to train the neural networks in these architectures as part of the system in an end-to-end manner.

A DeepProbLog program is a ProbLog program that is extended with a set of ground neural ADs (nADs) of the form
$nn(m_q, [X_1, ..., X_k], O, [y_1, ..., y_n]) :: q(X_1, ..., X_k,O)$. Here, $nn$ indicates that the following is an nAD and $m_q$ is a neural network identifier. The neural network $m_q$ will be provided the input vector $[X_1, ..., X_k]$ and output a probability distribution over the domain $O \in [y_1, ..., y_n]$. 
nADs work similarly to ADs in the sense that they provide a mutually-exclusive distribution of probabilities over a set of atoms. In nADs, however, these probabilities are generated from the output of a neural network, instead of being manually defined.
%More specifically, the parameters $[X_1, ..., X_k]$ are fed into a neural network, which then outputs a probability distribution for each of the classes $[y_1, ..., y_n]$. 
The sum of the probabilities over the domain $O$ must equal 1. In neural networks for multiclass classification, this is typically done by applying a softmax layer to real-valued output scores, a choice we also adopt in our experiments.
% For instance, if we wanted to classify MNIST digits into their respective values so that we can perform operations with them, we would specify the following nAD:

% \begin{verbatim}
% nn(mnist_net,[X],Y,[0,1,2,3,4,5,6,7,8,9]) :: digit(X,Y).
% \end{verbatim}

% This would then allow us to perform a query of the form $\mathtt{digit}(\digit{5},Y)$. This query would be transformed into an AD of the form 

% $$p_0::\mathtt{digit}(\digit{5},0);p_1::\mathtt{digit}(\digit{5},1);...;p_9::\mathtt{digit}(\digit{5},9)$$

% The values of the probabilities $p_i$ would be based on the output of the neural network. This neural network could take any shape, e.g., a convolutional network for image encoding, a recurrent network for sequence encoding, etc. However, its output layer, which feeds the corresponding neural predicate, needs to be normalized. In neural networks for multiclass classification, this is typically done by applying a softmax layer to real-valued output scores, a choice we also adopt in our experiments.

After defining the structure of the neural network and the logic level, it is possible to use DeepProbLog to infer the answers to our queries. To perform this inference, DeepProbLog transforms the logic layer into an arithmetic circuit and obtains the required probabilities from the neural network. This arithmetic circuit can then be used to calculate the probability that the query is true, based on the output of the neural network.
%The steps performed to obtain this arithmetic circuit are specific to each query. As such, they must be performed each time the system is sent a query.

In order to train the neural network, the system first performs inference as described above. Then, DeepProbLog is able to perform gradient-based learning. First, the arithmetic circuit used during the inference is also used to perform the gradient computations. Since this arithmetic circuit is composed of addition and multiplication operations, this means that it is differentiable. This allows DeepProbLog to compute the gradient with respect to the probabilistic logic program. This gradient can then be used to train the neural network using backpropagation. For a more detailed explanation on the technical aspects of DeepProblog's inference and learning, see \cite{MANHAEVE2021103504}.

%%%%%%%%%%%%%%%%%%%%%%%%%%%%%%%%%%%%%%%%%%%%%%%%%%%%%%%%%%%%%%%%%%%%%%%%%%%%%%%%%
%\section{Critical analysis of related work}
\section{Related work and its limitations}
\label{sec:critical_analysis}

In this section, we will explore the existing CEP approaches that are able to use subsymbolic data. We will also describe the limitations of each of those approaches, which our approach aims to solve.
There are three main types of approaches, further explained in the following sections: (i) using pre-trained neural networks to extract the symbolic information from the subsymbolic data, (ii) using a purely statistical approach and (iii) neuro-symbolic approaches. 
%In the following sections we will describe the limitations of each of those approaches, which our approach aims to solve.

\subsection{Pre-trained neural networks approaches}

Some CEP approaches use a pre-trained neural network to transform high-bandwidth data into symbolic information, allowing the user to define rules on it. For example, in \cite{ROLDAN2020113251} the authors show that this allows them to reduce the number of false positives in a system when detecting IoT security attacks. They use a neural network to predict the length of the suspected packets. If the predicted length does not match the actual length of the packet, a complex event is generated indicating that an attack might be happening. 
In \cite{Fusion}, we present a system that can detect different violent activities from a CCTV feed. A pre-trained neural network is used to process short segments of video (16 frames, about half a second) detecting potential violent acts. Another pre-trained neural network is used to detect people in the same video feed. A probabilistic logic program is then used to combine the outputs of these neural networks to detect the complex, violent events.
%These complex events are defined in a manner inspired by event calculus, based on the approach used in \cite{Skarlatidis2015}, identifying the conditions for a complex event to initiate and terminate. %By using ProbLog to define the rules for the complex events, the system is able to output a probability for each complex event to be happening at every point in time. This probability will depend on the confidence given in the inputs. In \cite{Fusion}, these probabilities came from the predictions made by the neural network.

%Both of these approaches are capable of using subsymbolic data as an input by using a neural network to extract the required information. This process transforms the subsymbolic data into symbolic data, which can then be used to define the rules to perform CEP. 

Both \cite{ROLDAN2020113251} and \cite{Fusion} use pre-trained neural networks to parse the simple events. In this paper, instead, we assume that no such pre-trained neural networks exist. As such, we assume that only end-to-end training is possible. This means that we only have training labels for when the complex events are happening, and not for the simple events. While this does make the training problem harder, it is undeniably easier to obtain labels for the complex events, thus reducing the costs associating to create the training set. %This is because complex events are the aggregation of simple events meaning that, by definition, less labels will be required for the same amount of training data. As such, being able to train using only end-to-end training would significantly reduce the cost of generating training data.

\subsection{Purely statistical approaches}
\label{sec:neural_only_approaches}

One possible approach is to view the whole CEP problem as a classification problem, and---for instance---use neural networks to detect when complex events occur. These approaches remove the manual definitions of complex events, and instead attempt to train the neural network to identify those definitions at the same time as it learns to classify the subsymbolic data. Due to the relevance of time in the definition of complex events, a Long Short Term Memory (LSTM) \cite{Mishra2018} or a Convolutional 3D layer (C3D) \cite{AAAI1817205} can be used. However, due to the necessity of learning the complex event rules, these approaches need very large amounts of data to train. Furthermore, the complexity of the rules that define the complex events is limited, due to the fact that the neural networks need to learn those rules. %As we will show in Section \ref{sec:results}, these approaches are able to learn to detect patterns of simple events if they are given enough training data. However, anything more complex would likely require infeasible amounts of training data.

\subsection{Neuro-symbolic approaches}
\label{sec:neuroplex}

The current state of the art in CEP with subsymbolic data is Neuroplex \cite{Neuroplex}. Neuroplex is a neuro-symbolic approach that makes use of human knowledge in order to reduce the amount of training data required when compared to purely statistical approaches. This is done by dividing the problem into two levels; low-level perception and high-level reasoning. The high-level reasoning is responsible for detecting the complex events based on manually defined rules, while the low-level perception is responsible for parsing the subsymbolic data into a set of classes that can be used when defining the rules.

In Neuroplex, the user defines the rules for the complex events. Then, a neural network is trained to emulate a logic layer that recognizes those rules. This allows users to inject human knowledge into the system. The neural network that emulates those rules is then used as the high-level reasoning. This is combined with another neural network, which performs the task of the low-level perception. Then, the high-level reasoning layer is frozen, meaning that the weights for this layer will not be modified by further training. Finally, the system is trained in an end-to-end manner. This trains the low-level perception neural network to recognize the simple events into the classes used to define the complex events.

Using a neural network to emulate the user defined rules is what allows Neuroplex to train in an end-to-end manner. However, it also introduces some limitations. Firstly, the reasoning neural network needs to be trained each time that the rules for the complex events are updated. As such, the whole system needs to be trained even if there only is a small change to the rules. Secondly, the ways in which complex events can be defined are, currently, substantially limited when compared to other CEP approaches. While improvements could be made to the system that trains the high-level reasoning to be more flexible, this would require a significant amount of work. At the moment, the high-level neural network can only be trained to recognize patterns of simple events within a given window. Finally, while Neuroplex can generate synthetic data to train the neural network to emulate the rules, it is not possible for the user to know if the neural network will behave exactly as the rules define in all situations. This is due to the nature of the neural network, which may give an unexpected answer if the given situation has not been seen in the training data. The only way to guarantee that the neural network will always behave as expected is to evaluate every possible situation, which becomes unfeasible as the complexity of the problem increases. 
%As a result, there is a risk Neuroplex will not be robust against some adversarial conditions. 
In this paper, we propose an architecture that aims to solve these issues.

%%%%%%%%%%%%%%%%%%%%%%%%%%%%%%%%%%%%%%%%%%%%%%%%%%%%%%%%%%%%%%%%%%%%%%%%%%%%%%%%%
\section{Neuro-symbolic processing of data streams}
\label{sec:methodology}

\begin{lstfloat}[]
\begin{lstlisting}
% Main interface for the framework.
sequence(S, W, T) :-
    reverse(S, S2), % Reverse list to simplify rule definitions
    sequenceEndingAt(S2, W, T).
% If the sequence is empty all events have happened
sequenceWithin([], _, _).
% S can be within W of T if the last element of S happens at T
sequenceWithin(S, W, T) :-
    sequenceEndingAt(S, W, T).
% S can be within W of T if it is within W-1 of T-1
sequenceWithin(S, W, T) :-
    W > 0, T >= 0,
    NextW is W - 1,
    allTimeStamps(Timestamps),
    previousTimeStamp(T, Timestamps, Tprev),
    sequenceWithin(S, NextW, Tprev).
% S will end at T if the first element of S (X) happens at T and the rest of the elements (L) happen within W-1 of T-1
sequenceEndingAt([X | L], W, T) :-
    W > 0, T >= 0,
    digit(T, X),
    NextW is W - 1,
    allTimeStamps(Timestamps),
    previousTimeStamp(T, Timestamps, Tprev),
    sequenceWithin(L, NextW, Tprev).
\end{lstlisting}
\caption{Sequence framework.}
\label{lst:sequence_framework}
\end{lstfloat}

\newcommand{\nndraw}[1]{%
\def\layersep{2.5cm}
\begin{tikzpicture}[node distance=\layersep,scale={#1}, every node/.style={scale={#1}}]
    %\tikzstyle{every pin edge}=[<-,shorten <=1pt]
    \tikzstyle{neuron}=[circle,minimum size=0.2pt,inner sep=0pt,line width=0.05mm,draw,scale=0.3]
    \tikzstyle{input neuron}=[neuron]%, fill=black];
    \tikzstyle{output neuron}=[neuron]%, fill=black];
    \tikzstyle{hidden neuron}=[neuron]%, fill=black];
    %\tikzstyle{annot} = [text width=4em, text centered]

    % Draw the input layer nodes
    \foreach \name / \y in {1,...,4}
    % This is the same as writing \foreach \name / \y in {1/1,2/2,3/3,4/4}
        \node[input neuron] (I-\name) at (0,-\y) {};

    % Draw the hidden layer nodes
    \foreach \name / \y in {1,...,5}
        \path[yshift=0.5cm]
            node[hidden neuron] (H-\name) at (\layersep,-\y cm) {};

    % Draw the output layer node
    \foreach \name / \y in {1/2,2/3,3/4}
        \path[yshift=0.5cm]
            node[output neuron] (O-\name) at (2*\layersep,-\y cm) {};
    % \node[output neuron,pin={[pin edge={->}]right:Output}, right of=H-3] (O) {};

    % Connect every node in the input layer with every node in the
    % hidden layer.
    \foreach \source in {1,...,4}
        \foreach \dest in {1,...,5}
            \path (I-\source) edge (H-\dest);

    % Connect every node in the hidden layer with the output layer
    \foreach \source in {1,...,5}
        \foreach \dest in {1,...,3}
            \path (H-\source) edge (O-\dest);

    % % Annotate the layers
    % \node[annot,above of=H-1, node distance=1cm] (hl) {Hidden layer};
    % \node[annot,left of=hl] {Input layer};
    % \node[annot,right of=hl] {Output layer};
\end{tikzpicture}}

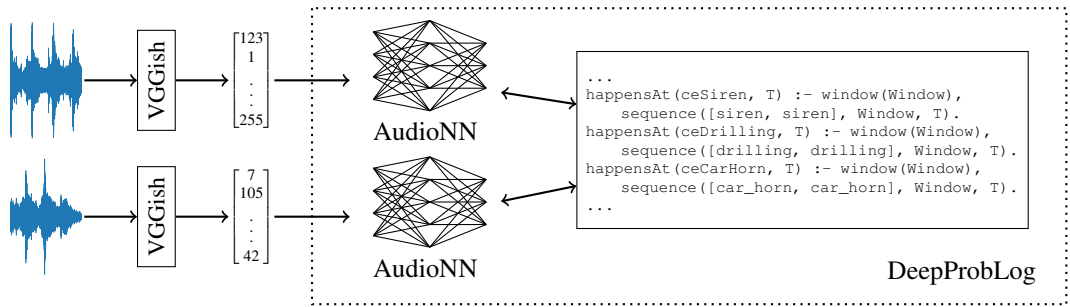
\begin{figure*}
    \centering

    \begin{tikzpicture}[scale=1, transform shape]
    \node[inner sep=0pt] (audio1) {\includegraphics[height=1.8cm, width=1cm]{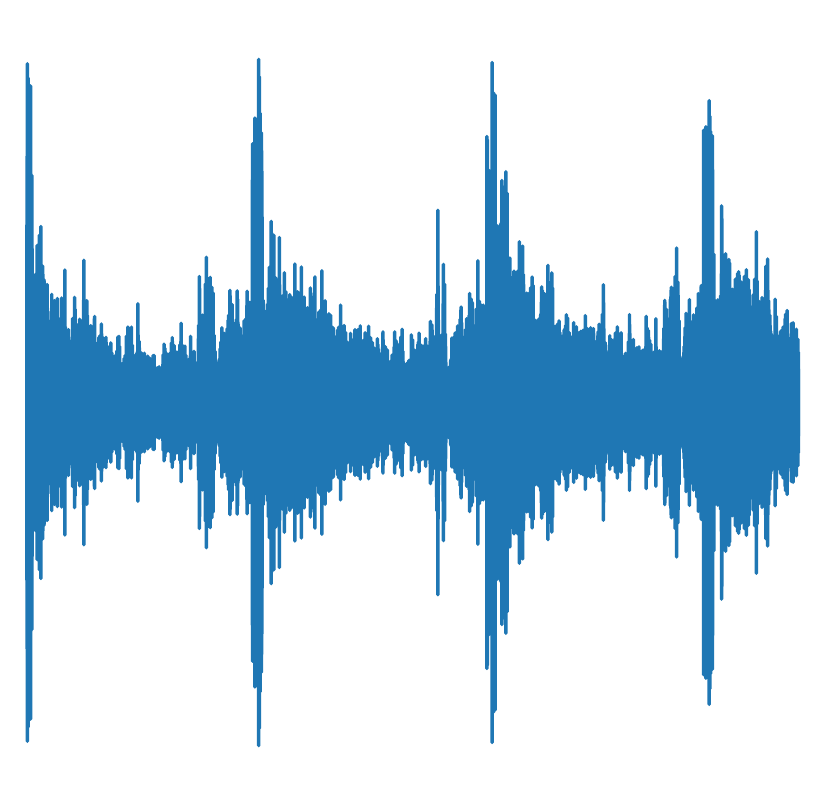}};
    \node[inner sep=4pt, right=0.7cm of audio1, rectangle, draw] (VGGish1) {\rotatebox{90}{\footnotesize VGGish}};
    \node[inner sep=0pt, below=0cm of audio1] (audio2) {\includegraphics[height=1.8cm, width=1cm]{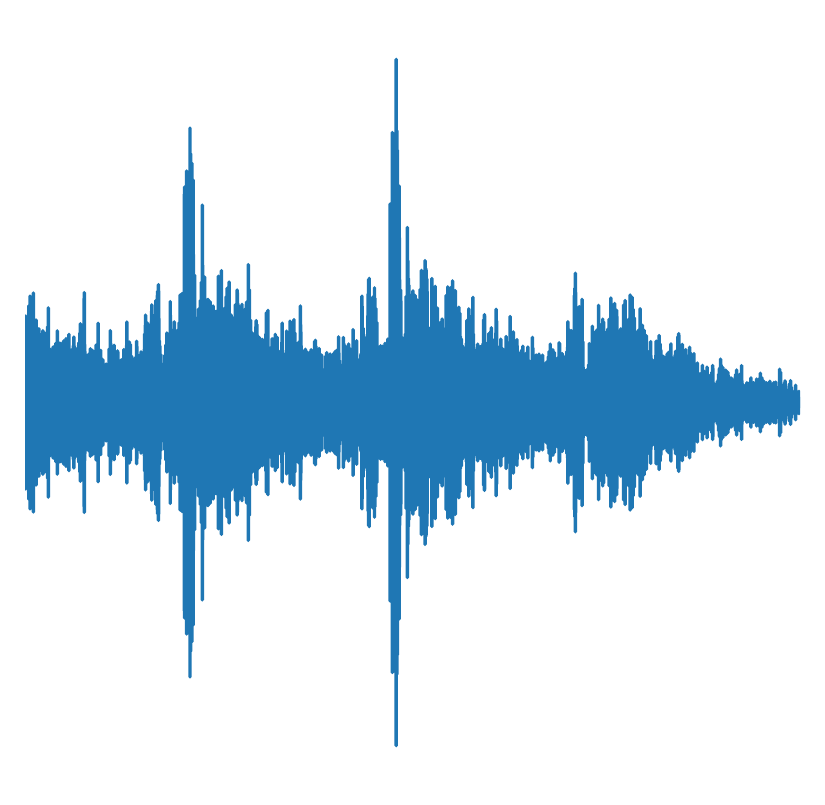}};
    \node[inner sep=4pt, right=0.7cm of audio2, rectangle, draw] (VGGish2) {\rotatebox{90}{\footnotesize VGGish}};
    \node[inner sep=0pt, right=0.7cm of VGGish1] (audio1tensor) {\tiny
        $\begin{bmatrix}
        123\\
        1\\
        \vdots\\
        255
        \end{bmatrix}$
        };
    \node[inner sep=0pt, right=0.7cm of VGGish2] (audio2tensor) {\tiny
        $\begin{bmatrix}
        7\\
        105\\
        \vdots\\
        42
        \end{bmatrix}$
        };
    \node[inner sep=0pt,right=1cm of audio2tensor, text width=2cm, align=center] (nnaudio2) {\nndraw{0.3} AudioNN};
    \node[inner sep=0pt,right=1cm of audio1tensor, text width=2cm, align=center] (nnaudio1) {\nndraw{0.3} AudioNN};
    \node[inner sep=0pt,below right=-1.2cm and 1cm of nnaudio1,rectangle,draw, text width=6cm] (problogaudio) {
     \tiny
     \begin{verbatim}
 ...
 happensAt(ceSiren, T) :- window(Window),
     sequence([siren, siren], Window, T).
 happensAt(ceDrilling, T) :- window(Window),
     sequence([drilling, drilling], Window, T).
 happensAt(ceCarHorn, T) :- window(Window),
     sequence([car_horn, car_horn], Window, T).
 ...
     \end{verbatim}
     };
    \draw[->,thick] (audio1) -- (VGGish1);
    \draw[->,thick] (VGGish1) -- (audio1tensor);
    \draw[->,thick] (audio2) -- (VGGish2);
    \draw[->,thick] (VGGish2) -- (audio2tensor);
    \draw[->,thick] (audio1tensor.east) -- (nnaudio1);
    \draw[->,thick] (audio2tensor.east) -- (nnaudio2);
    \draw[<->,thick] (nnaudio1) -- (problogaudio);
    \draw[<->,thick] (nnaudio2) -- (problogaudio);
    
    \draw[thick,dotted]     ($(nnaudio1.north west)+(-0.5,0.15)$) rectangle ($(problogaudio.south east)+(0.5,-1.0)$);
    
    \node[below right=0.3cm and -2cm of problogaudio] {DeepProbLog};
    \end{tikzpicture}

    \caption{Overall architecture of of our approach for the experiments performed in this paper.}
    \label{fig:overall-architecture}
\end{figure*}

In this section, we describe how we have used DeepProbLog to implement a neuro-symbolic approach to CEP. Our approach allows users to inject human knowledge into the system by manually defining rules for the complex events. At the same time, it allows us to perform end-to-end training in order to make use of subsymbolic data such as audio. This is archived by dividing the tasks into two distinct levels; (i) a perception level, where a neural network is used to classify subsymbolic data in order to extract the symbolic information and (ii) a reasoning level, where probabilistic logic programming is used to define the complex event rules.

As explained above in Section \ref{sec:neuroplex}, Neuroplex \cite{Neuroplex} also divides the problem into perception and reasoning levels. However Neuroplex uses a neural network to emulate the rules, instead of using an explicit logic layer. By using an explicit logic layer in our approach we remove the need of training a neural network to emulate the functionality of the logic layer, which makes it easier to update the complex event rules. Furthermore, we also remove the risk of the neural network behaving in an unexpected manner, thus providing a higher robustness.

In this paper, we use audio files as an input to the system. For processing purposes, the input audio is divided into one second segments, each of which is considered a simple event. The system then tries to detect the complex events defined by the user in the input stream. The user can define what constitutes a complex event using ProbLog. For this paper, we have used the clause \textit{sequence}, which will be true if a given sequence of simple events $S$ happens within a given window $W$ at a specific timestamp $T$, with the last element of $S$ happening at $T$ and all other elements of $S$ happening in the right order between $T-W$ and $T$. Listing \ref{lst:sequence_framework} shows the framework that defines the clause \textit{sequence}.

%Then, we try to detect patterns where a sound of the same class occurs twice within the given window. More specifically, we look for cases in which the same class occurs at the last position in the window size and at some other position within the window. When this happens, a complex event is generated. The type of complex event will depend on the class of the repeated simple event.

Figure \ref{fig:overall-architecture} shows the diagram used for our experimentation. Firstly, the input audio is divided into one second segments and pre-processed. For this, we use VGGish \cite{hershey2017cnn}, a state-of-the-art feature extractor for audio classification models\footnote{In order to make it compatible with DeepProbLog, we use a PyTorch implementation of VGGish, available at \url{https://github.com/harritaylor/torchvggish}}. VGGish performs a feature extraction process which results in a matrix of size $128 \times N$, where $N$ is the length of the input audio file in seconds. Each position in the matrix contains a vaule between 1 and 255. After performing this pre-processing, the resulting matrix is fed into our system. The vector resulting from each 1 second segment is fed into a multilayer perceptron (MLP) neural network, AudioNN in the diagram. This neural network classifies the segment into one of the 10 classes that appear in our dataset. The MLP used in our experimentation has 5 layers with 100, 80, 50, 25 and 10 neurons, in this order. A ReLU activation function is used between each of the layers, and a Softmax activation function is applied at the end.

Finally, the logic layer makes use of the output values from the neural network to predict whether or not a certain complex event is happening at a certain point in time. In order to determine this, the rules provided by the user are used. The diagram also shows a snippet of the logic rules used to define the complex events. This code defines that the complex events happen if a specific pattern of simple events happens within a given window of time, using the clause \textit{sequence} as defined in Listing \ref{lst:sequence_framework}. For the full code, see \url{https://github.com/dais-ita/DeepProbCEP}.

For the experimentation in this paper, we set a maximum number of epochs of 100. However, in order to avoid overfitting we also make use of early stopping with a patience of 10 epochs. This means that if the performance of the system on the validation dataset does not improve for 10 epochs, we end the training early. We will then use the weights that performed the best in the validation dataset for testing.

%%%%%%%%%%%%%%%%%%%%%%%%%%%%%%%%%%%%%%%%%%%%%%%%%%%%%%%%%%%%%%%%%%%%%%%%%%%%%%%%%
\section{Datasets generation}
\label{sec:generating_datasets}

\begin{figure*}
    \centering
    \begin{tikzpicture}[scale=0.9, transform shape]
        \def\inhsep{0.5cm}
        \def\shufflehsep{2.2cm}
        \def\hsep{0.8cm}
        \def\vsep{0.0cm}
        \def\traininghsep{0.1cm}
        \def\mw{1.3cm}
        \def\mwClass{2.5cm}
        \def\mh{20pt}
        
        \def\myfontsize{\footnotesize}
        
        \def\cecolor{red}
        
        \tikzstyle{titleStyle}=[text centered, minimum height=\mh, font=\myfontsize]
        \tikzstyle{audioStyle}=[text centered, minimum height=\mh, font=\myfontsize]
        \tikzstyle{classStyle}=[text centered, minimum height=\mh, font=\myfontsize, minimum width=\mwClass]
        \tikzstyle{timestampStyle}=[text centered, minimum height=\mh, font=\myfontsize]
        \tikzstyle{ceClassStyle}=[text centered, minimum height=\mh, font=\myfontsize, minimum width=\mw]
        \tikzstyle{rectangleStyle}=[thick, dotted, inner xsep=2mm, inner ysep=0mm]

        \node[audioStyle] (audioImage0) {\audioImage{4}};
        \node[audioStyle, below=\vsep of audioImage0] (audioImage1) {\audioImage{9}};
        \node[audioStyle, below=\vsep of audioImage1] (audioImage2) {\audioImage{5}};
        \node[audioStyle, below=\vsep of audioImage2] (audioImage3) {\audioImage{0}};
        \node[audioStyle, below=\vsep of audioImage3] (audioImage4) {\audioImage{4}};
        \node[audioStyle, below=\vsep of audioImage4] (audioImage5) {\audioImage{6}};
        \node[audioStyle, below=\vsep of audioImage5] (audioImage6) {\audioImage{7}};
        \node[audioStyle, below=\vsep of audioImage6] (audioImage7) {\audioImage{2}};
%        \node[audioStyle, below=\vsep of audioImage7] (audioImage8) {\audioImage{6}};

        \node[titleStyle, above=\vsep of audioImage0] (audioImage) {Audio};

        \node[classStyle, right=\inhsep of audioImage0] (audioClass0) {siren};
        \node[classStyle, right=\inhsep of audioImage1] (audioClass1) {street\_music};
        \node[classStyle, right=\inhsep of audioImage2] (audioClass2) {drilling};
        \node[classStyle, right=\inhsep of audioImage3] (audioClass3) {air\_conditioner};
        \node[classStyle, right=\inhsep of audioImage4] (audioClass4) {siren};
        \node[classStyle, right=\inhsep of audioImage5] (audioClass5) {enginge\_idling};
        \node[classStyle, right=\inhsep of audioImage6] (audioClass6) {gun\_shot};
        \node[classStyle, right=\inhsep of audioImage7] (audioClass7) {children\_playing};
%        \node[classStyle, right=\inhsep of audioImage8] (audioClass8) {enginge\_idling};

        \node[titleStyle, above=\vsep of audioClass0] (audioClass) {Class};

        \node (auxAudio) at ($(audioImage)!0.5!(audioClass)$) {};
        \node[titleStyle, above=\vsep of auxAudio] (audio) {Urban Sounds 8K};
        
        \node[draw, rectangleStyle, fit=(audio) (audioImage) (audioClass) (audioImage7) (audioClass7)] (audioRectangle) {};

        \node[timestampStyle, right=\shufflehsep of audioClass0] (timestamp0) {0};
        \node[timestampStyle, right=\shufflehsep of audioClass1] (timestamp1) {1};
        \node[timestampStyle, right=\shufflehsep of audioClass2] (timestamp2) {2};
        \node[timestampStyle, right=\shufflehsep of audioClass3] (timestamp3) {3};
        \node[timestampStyle, right=\shufflehsep of audioClass4] (timestamp4) {4};
        \node[timestampStyle, right=\shufflehsep of audioClass5] (timestamp5) {5};
        \node[timestampStyle, right=\shufflehsep of audioClass6] (timestamp6) {6};
        \node[timestampStyle, right=\shufflehsep of audioClass7] (timestamp7) {7};
%        \node[timestampStyle, right=\shufflehsep of audioClass8] (timestamp8) {8};
        
        \node[titleStyle, above=\vsep of timestamp0] (sTimestamp) {Timestamp};
        
        \node[audioStyle, right=\hsep of timestamp0] (sImage0) {\audioImage{0}};
        \node[audioStyle, right=\hsep of timestamp1] (sImage1) {\audioImage{7}};
        \node[audioStyle, right=\hsep of timestamp2] (sImage2) {\audioImage{6}};
        \node[audioStyle, right=\hsep of timestamp3] (sImage3) {\audioImage{4}};
        \node[audioStyle, right=\hsep of timestamp4] (sImage4) {\audioImage{5}};
        \node[audioStyle, right=\hsep of timestamp5] (sImage5) {\audioImage{4}};
        \node[audioStyle, right=\hsep of timestamp6] (sImage6) {\audioImage{2}};
%        \node[audioStyle, right=\hsep of timestamp7] (sImage7) {\audioImage{6}};
%        \node[audioStyle, right=\hsep of timestamp8] (sImage8) {\audioImage{9}};
        \node[audioStyle, right=\hsep of timestamp7] (sImage7) {\audioImage{9}};
        
        \node[titleStyle, above=\vsep of sImage0] (sImage) {Audio};

        \node[classStyle, right=\inhsep of sImage0] (sClass0) {air\_conditioner};
        \node[classStyle, right=\inhsep of sImage1] (sClass1) {gun\_shot};
        \node[classStyle, right=\inhsep of sImage2] (sClass2) {enginge\_idling};
        \node[classStyle, right=\inhsep of sImage3, \cecolor] (sClass3) {siren};
        \node[classStyle, right=\inhsep of sImage4] (sClass4) {drilling};
        \node[classStyle, right=\inhsep of sImage5, \cecolor] (sClass5) {siren};
        \node[classStyle, right=\inhsep of sImage6] (sClass6) {children\_playing};
%        \node[classStyle, right=\inhsep of sImage7] (sClass7) {enginge\_idling};
%        \node[classStyle, right=\inhsep of sImage8] (sClass8) {street\_music};
        \node[classStyle, right=\inhsep of sImage7] (sClass7) {street\_music};

        \node[titleStyle, above=\vsep of sClass0] (sClass) {Class};

        \node (auxS) at ($(sTimestamp)!0.5!(sClass)$) {};
        \node[titleStyle, above=\vsep of auxS] (s) {$S$};
        
        \node[draw, rectangleStyle, \cecolor, fit=(timestamp1) (sClass5)] (ce) {};

        \node[draw, rectangleStyle, fit=(s) (sTimestamp) (sClass) (timestamp7) (sClass7) (ce)] (sRectangle) {};
        
        \node[ceClassStyle, right=\hsep of sClass0, minimum width=\mw] (c0) {Null};
        \node[ceClassStyle, right=\hsep of sClass1, minimum width=\mw] (c1) {Null};
        \node[ceClassStyle, right=\hsep of sClass2, minimum width=\mw] (c2) {Null};
        \node[ceClassStyle, right=\hsep of sClass3, minimum width=\mw] (c3) {Null};
        \node[ceClassStyle, right=\hsep of sClass4, minimum width=\mw] (c4) {Null};
        \node[ceClassStyle, right=\hsep of sClass5, minimum width=\mw, \cecolor] (c5) {ceSiren};
        \node[ceClassStyle, right=\hsep of sClass6, minimum width=\mw] (c6) {Null};
        \node[ceClassStyle, right=\hsep of sClass7, minimum width=\mw] (c7) {Null};
%        \node[ceClassStyle, right=\hsep of sClass8, minimum width=\mw] (c8) {Null};

        \node[titleStyle, above=\vsep of c0] (c) {$C$};
        
        \node[draw, rectangleStyle, fit=(c) (c7)] (cRectangle) {};

        \node[audioStyle, right=\hsep of c0] (ts0) {\audioImage{0}};
        \node[audioStyle, right=\hsep of c1] (ts1) {\audioImage{7}};
        \node[audioStyle, right=\hsep of c2] (ts2) {\audioImage{6}};
        \node[audioStyle, right=\hsep of c3] (ts3) {\audioImage{4}};
        \node[audioStyle, right=\hsep of c4] (ts4) {\audioImage{5}};
        \node[audioStyle, right=\hsep of c5] (ts5) {\audioImage{4}};
        \node[audioStyle, right=\hsep of c6] (ts6) {\audioImage{2}};
%        \node[audioStyle, right=\hsep of c7] (ts7) {\audioImage{6}};
%        \node[audioStyle, right=\hsep of c8] (ts8) {\audioImage{9}};
        \node[audioStyle, right=\hsep of c7] (ts7) {\audioImage{9}};

        \node[titleStyle, above=\vsep of ts0] (ts) {$TS$};
        
        \node[ceClassStyle, right=\traininghsep of ts0, minimum width=\mw] (tc0) {Null};
        \node[ceClassStyle, right=\traininghsep of ts1, minimum width=\mw] (tc1) {Null};
        \node[ceClassStyle, right=\traininghsep of ts2, minimum width=\mw] (tc2) {Null};
        \node[ceClassStyle, right=\traininghsep of ts3, minimum width=\mw] (tc3) {Null};
        \node[ceClassStyle, right=\traininghsep of ts4, minimum width=\mw] (tc4) {Null};
        \node[ceClassStyle, right=\traininghsep of ts5, minimum width=\mw] (tc5) {ceSiren};
        \node[ceClassStyle, right=\traininghsep of ts6, minimum width=\mw] (tc6) {Null};
        \node[ceClassStyle, right=\traininghsep of ts7, minimum width=\mw] (tc7) {Null};
%        \node[ceClassStyle, right=\traininghsep of ts8, minimum width=\mw] (tc8) {Null};

        \node[titleStyle, above=\vsep of tc0] (tc) {$C$};
        
        \node (auxTraining) at ($(ts)!0.5!(tc)$) {};
        \node[titleStyle, above=\vsep of auxTraining] (training) {Training};
        
        \node[draw, rectangleStyle, fit=(training) (ts) (tc) (ts7) (tc7)] (trainingRectangle) {};
        
        \draw[->,thick, font=\myfontsize] (audioRectangle) -- node[anchor=south] {Shuffle} (sRectangle);
        \draw[->, thick, \cecolor, dotted] (sClass5.east -| ce.east) -- (c5);

        %\draw[decorate, decoration={brace, amplitude=10pt, raise=20pt}, yshift=0pt] (sClass2.north east) -- (sClass5.south east);
        \draw[->, thick] (trainingRectangle.west -| cRectangle.east) -- (trainingRectangle.west);
    \end{tikzpicture}
    
    \caption{Diagram representing how the datasets used in this paper are generated. A window of 5 has been used. After randomly shuffling the audio-class pairs from the Urban Sounds 8K dataset, we detect that at timestamp 5 we have two instances of the class \textit{siren} within the given window. Therefore, we mark timestamp 5 in $C$ as $ceSiren$. We can also observe that there is a pair of \textit{engine\_idling} on timestamps 2 and 7. However, the distance between them is bigger than the given window, and therefore that does not result in a complex event. Finally, for the training dataset we remove the ground truth for the sound class, as we are doing end-to-end training.}
    \label{fig:dataset_generation}
\end{figure*}
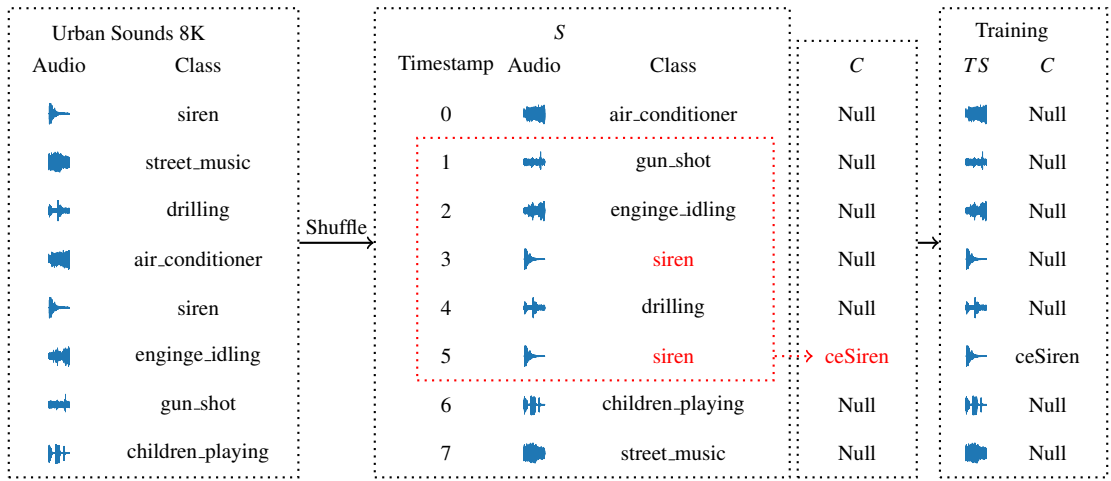

In this section, we describe how we have generated the datasets used to evaluate our approach. All the datasets are generated using Urban Sounds 8K \cite{UrbanSound}, a dataset containing over eight thousand short audio files (4 seconds or less) that contain sounds from 10 different classes: air\_conditioner, car\_horn, children\_playing, dog\_bark, drilling, enginge\_idling, gun\_shot, jackhammer, siren, and street\_music.

The first dataset used in our experiments are the \textit{base datasets}. These datasets allow us to evaluate how the size of the sliding window affects the performance of the system. In our approach, we use this sliding window to define the maximum amount of time that can pass between the first and last simple events that will be aggregated into a complex event. As such, if a set of simple events follow the pattern we have defined but they are too far apart in a temporal sense, no complex event will be generated. This allows us to define that simple events that are separated by large amounts of time have no relation to each other.

We also want to evaluate how robust or approach is, as defined in the fourth objective from Section \ref{sec:introduction}. For this purpose, we have generated datasets where a percentage of the training labels have been randomly changed. Different percentage values are used to evaluate how this affects our approach. We call this dataset type \textit{random noise dataset}.

In the following sections we will give more details on how both types of datasets have been generated.

For both types of datasets, we are using the same definitions for the complex events. Specifically, we are looking for patterns in which the sound that occurs in the last position of our sliding window also appears in another position within the window size. Each of the 10 sound classes in Urban Sounds 8K generates a different class of complex event.

\subsection{Base dataset}
\label{sec:base_dataset}

In this section, we describe how we generated the base dataset. The process used to generate the base dataset allows us to change the window size by changing the value of $Window$. $Window$ is a positive integer that indicates the number of timestamps between the first and last simple events that form a complex event. For this paper, we have generated datasets with window sizes of 2, 3, 4 and 5.

In order to generate the base dataset, we use the different folds from Urban Sounds 8K. Out of the 10 folds provided by the original dataset, 8 are used to generate our training dataset, 1 is used to generate our validation dataset and the last fold is used to generate our testing dataset. The steps to generate the base datasets are shown in Figure \ref{fig:dataset_generation}, which illustrates the following steps:

\begin{enumerate}
    \item We take all the audio files from the original dataset and randomly shuffle them into a sequence $S$ of simple events, where each audio file represents one simple event. Therefore, the length of $S$ is the number of audio files in the original dataset. Simple events can be accessed by their index like so $S[I]$. For each of them we can access the file itself and the class it contains using $S[I].audio$ and $S[I].class$, respectively. In order to have a consistent length for all simple events, only the first second of each audio file is used.
    
    \item We create a list $C$ that will indicate for each timestamp whether a complex event happens. We initialize this list with \textit{null}, which hereinafter represents that no complex event happens at the specified timestamp.
    
    \item For each timestamp $T$ where $0 < T < len(S)$:
    
    \begin{enumerate}
        \item \label{step:labelling_ce} If the pattern for one of the complex events occurs, mark $C[T]$ as the corresponding complex event. Formally, if there exists $P$ such that $T - Window < P \leq T$ and $S[P].class = S[T].class$, mark $C[T]$ as $ceN$, where $N$ is the value of $S[T].class$. This means that if a sound occurs at the last position in the window $T$ and somewhere else within the window $P$, a complex event is generated.
        
        \item Otherwise, leave $C[T]$ marked as the null class.
    \end{enumerate}
    
    \item Finally, if this is the training dataset, generate the training sequence of simple events $TS$, which will only contain the audio files, but not the ground truth of which class they represent, as these should not be available when performing end-to-end training. Therefore, $TS[I] = S[I].audio$ for $0 < I < len(S)$.
\end{enumerate}

Before using these datasets for training, they are also balanced in order to avoid overfitting for a specific class. This results in a training dataset with 1000 training points for each window size.

\subsection{Random noise datasets}
\label{sec:generating_datasets:random_noise}

Given the complexity of the definition of some of the real world complex events, it can sometimes be hard to correctly label when a certain complex event is happening. This can lead to errors on the training dataset, which might affect the accuracy and confidence of the system after training. This can also happen is due to a malicious attack that is intended to reduce the performance of our system. In Section \ref{sec:introduction}, we defined that one of our objectives was to be robust against noisy data in our training datasets.
Of course, this is not an issue with our synthetically generated dataset. However, using a synthetically generated dataset offers us the opportunity of artificially introducing noise in a controlled manner. This allows us to evaluate how well our approach might perform when used on a real dataset, which might contain an unknown level of noise.

To evaluate how robust our approach is, we have created datasets with different amounts of noise by randomly changing different percentages of the training labels for another random label, which simulates this noise. The noisy part of the dataset will have randomly assigned training labels, instead of the ones that should be assigned according to the complex event rules. We have generated datasets with \textit{percentages of noise} between 0.0 and 0.6, with a step of 0.2. For this, 0.0 means that no noise has been introduced to the dataset, while 1.0 would mean that the whole dataset consists of noisy labels. 

%For instance, assume that, based on the complex event definitions, a certain timestamp is marked as $ceSirens$. However, if this case is affected by the noise it will be labelled as another complex event, which is selected randomly every time. We will call the datasets generated using this process \textit{random noise datasets}.
In order to generate the random noise datasets, we use the same steps described to generate the base dataset, explained above in Section \ref{sec:base_dataset}. However, each time we label a timestamp as a complex event (Step \ref{step:labelling_ce}), there is probability that the label will become noisy. If that happens, instead of assigning the correct label according to the rules, a randomly chosen complex event label will be assigned. Note that this can happen irrespectively of what the original complex event label would have been. This probability is determined by the intended percentage of noise in the dataset, as defined above. Finally, the datasets are balanced. These datasets also have a size of 1000 training points.

It is important to note that this attack is only performed on the training dataset. This means that the testing dataset will maintain the ground truth, which will allow us to see how the system would perform in a real life scenario.

This kind of noise might appear both due to a malicious agent, and to the difficulty of labelling the sophisticated scenarios where a CEP system would be useful: for instance, different annotators might be having different consideration on what constitutes the a complex event. %It might also be due to incorrect labelling by the annotators where they select the incorrect class by accident.

%%%%%%%%%%%%%%%%%%%%%%%%%%%%%%%%%%%%%%%%%%%%%%%%%%%%%%%%%%%%%%%%%%%%%%%%%%%%%%%%%
\section{Experimental analysis}
\label{sec:results}

In this section, we explore the accuracy results for our approach after training with the synthetic datasets explained in Section \ref{sec:generating_datasets}.
All the values displayed on the graphs and tables in the following sections are the result of averaging the accuracies of 3 different executions.

\subsection{Performance with base dataset}
\label{sec:results:audio_setting}

\begin{table}[]
\centering
\caption{Average accuracy results and standard deviation for complex events classification by window size.
% These values are the average from 5 executions. The standard deviations across the 5 executions are also shown in the table (Sound STD and Pattern STD, respectively). Marked in bold the best performance for each window in terms of individual accuracy and pattern accuracy.
}
\label{tbl:audio_results}
\begin{tabular}{c c c}
\toprule
\textbf{Window size} & \textbf{Accuracy} & \textbf{STD} \\
\midrule
\textbf{2}	& 0.8657	& 0.0041	\\
\textbf{3}	& 0.7645	& 0.0109	\\
\textbf{4}	& 0.7069	& 0.0191	\\
\textbf{5}	& 0.6401	& 0.0225	\\
\bottomrule
\end{tabular}%
\end{table}

In Table \ref{tbl:audio_results} we can see the results of training our approach on a balanced dataset with 1000 training data points. As shown in the table, the performance of the approach is fairly good with a window size of 2. However, the performance does decrease as the window size increases. This could have been expected, as the problem gets more complex as the window size increases. This is because a bigger window size contains more simple events, which makes it more likely that the system will incorrectly classify one of them. This can cause the system to predict that a complex event is happening when it is not, thus reducing the performance of the system.

\subsection{Robustness against random noise}
\label{sec:results:random_noise}

\begin{figure}[]
    \centering
    \includegraphics[width=.9\columnwidth]{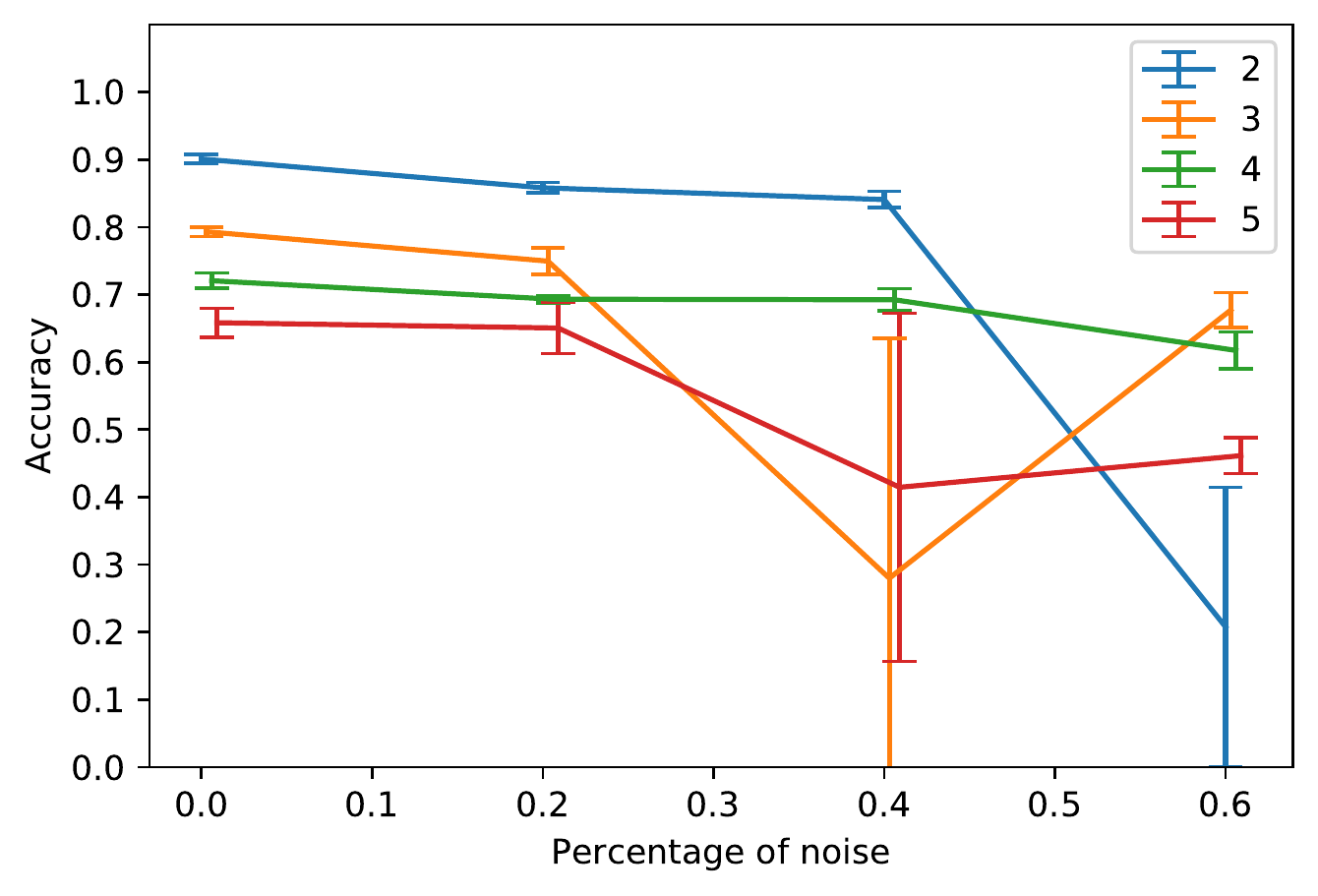}
    \caption{Evaluation of system's performance with different levels of noise in the training data. The horizontal axis indicates the percentage of training data points where the class has been randomly selected. The vertical axis indicates the accuracy of the system for either individual classifications or for complex event detection. The error bars represent the standard deviation for each case. The error bars have been slightly offset in their horizontal axis to avoid overlapping, which would make them hard to read.}
    \label{fig:scenario103}
\end{figure}

As explained above, we also want to know how robust our approach is against noisy training data. For this purpose, we have trained the system with the random noise datasets (explained above in Section \ref{sec:generating_datasets:random_noise}) and evaluated how the performance is affected. The results are shown in Figure \ref{fig:scenario103}. As seen in the graph, while there is a slight decrease in performance when training with the 20\% noise dataset, it does not seem to significantly impact the system. By contrast, with a percentage of noise of 40\% or higher, the performance seems much less consistent. While, in some cases, the system is still able to train correctly, in others it performs very significantly worse. This is what causes the high standard deviation that can be seen in the graph. As such, we cannot consider the system reliable under those percentages of noise. However, it does not seem likely that the user would not realize that almost half of the training points in the dataset are incorrectly labelled. As such, we would argue that our approach is robust against moderate amounts of noise.

%Also as expected, with a noise percentage of 1.0 the system obtains a very bad accuracy, with a performance comparable to a random classifier. This is because, in this situation, all of the data is noise, which means that it is impossible to learn anything from it.

%%%%%%%%%%%%%%%%%%%%%%%%%%%%%%%%%%%%%%%%%%%%%%%%%%%%%%%%%%%%%%%%%%%%%%%%%%%%%%%%%
\section{Conclusion and future work}
\label{sec:conclusion}

In this paper we have presented a neuro-symbolic approach capable of performing CEP on subsymbolic data. More specifically, we have demonstrated that our approach is capable of detecting complex events from an audio stream after training end-to-end. We have also shown that our approach is robust against a moderate amount of noise in the training data, thus fulfilling all four objectives defined in Section \ref{sec:introduction}.

As part of future work, we are considering on evaluating the performance on other types of subsymbolic data. This should be possible using a neural network architecture capable of classifying that type of subymbolic data.

An other area on which future research could be applied is on the time efficiency of the approach. Due to the use of a logic layer, our approach is slower in both training time and inference time when compared to approaches that are implemented using a neural network, such as Neuroplex \cite{Neuroplex}. This is mostly due to the cost of generating the arithmetic circuit used to calculate the output for the logic layer. DeepProbLog offers a cache functionality to reduce the amount of times this arithmetic circuit has to be generated. However, some further research will be needed to make the most out of this functionality for problems that deal with a temporal aspect.

\section{Acknowledgements}

This research was sponsored by the U.S. Army Research Laboratory and the U.K. Ministry of Defence under Agreement Number W911NF-16-3-0001. The views and conclusions contained in this document are those of the authors and should not be interpreted as representing the official policies, either expressed or implied, of the U.S. Army Research Laboratory, the U.S. Government, the U.K. Ministry of Defence or the U.K. Government. The U.S. and U.K. Governments are authorized to reproduce and distribute reprints for Government purposes notwithstanding any copyright notation hereon.

\bibliography{biblio.bib}
\bibliographystyle{aaai}

\appendix

\end{document}